\documentstyle{article}

\begin{document}

\title{Scar functions in the Bunimovich Stadium billiard.}

\author{Gabriel G. Carlo(1), Eduardo G. Vergini(2) \\
 and Pablo Lustemberg(2) \\
(1)Max-Planck-Institut f\"ur Physik komplexer Systeme,\\ 
N\"othnitzer Str. 38, D-01187 Dresden, Germany.\\
(2)Departamento de F{\'\i}sica, Comisi{\'o}n Nacional de \\
Energ{\' \i}a At{\'o}mica. Av. del Libertador 8250, \\
1429 Buenos Aires, Argentina.}

\date{\today}
\maketitle

\begin{abstract}
In the context of the semiclassical theory of short periodic orbits, 
scar functions play a crucial role. These wavefunctions live in the 
neighbourhood of the trajectories, resembling the hyperbolic 
structure of the phase space in their immediate vicinity. This property 
makes them extremely suitable for investigating chaotic eigenfunctions. 
On the other hand, for all practical purposes reductions to 
Poincar\'e sections become essential. Here we give a detailed 
explanation of resonances and scar functions construction in the 
Bunimovich stadium billiard and 
the corresponding reduction to the boundary. Moreover, we develop a 
method that takes into account the departure of the unstable and stable 
manifolds from the linear regime. This new feature extends the validity 
of the expressions.
\end{abstract}
\newpage

%
%

%
\def\temp{1.31}
\let\tempp=\relax
\expandafter\ifx\csname psboxversion\endcsname\relax
  \message{version: \temp}
\else
    \ifdim\temp cm>\psboxversion cm
      \message{version: \temp}
    \else
      \message{psbox(\psboxversion) is already loaded: I won't load
        psbox(\temp)!}
      \let\temp=\psboxversion
      \let\tempp= 
    \fi
\fi
\tempp
\let\psboxversion=\temp
\catcode`\@=11
%
%
\def\execute#1{#1}
\def\psm@keother#1{\catcode`#112\relax}
\def\executeinspecs#1{%
\execute{\begingroup\let\do\psm@keother\dospecials\catcode`\^^M=9#1\endgroup}}
%
%
\def\psfortextures{
\def\PSspeci@l##1##2{%
\special{illustration ##1\space scaled ##2}%
}}
\def\psfordvitops{
\def\PSspeci@l##1##2{%
\special{dvitops: import ##1\space \the\drawingwd \the\drawinght}%
}}
\def\psfordvips{
\def\PSspeci@l##1##2{%
\d@my=0.1bp \d@mx=\drawingwd \divide\d@mx by\d@my%
\includegraphics{##1\space}%
}}
\def\psforoztex{
\def\PSspeci@l##1##2{%
\special{##1 \space
      ##2 1000 div dup scale
      \putsp@ce{\number-\psllx} \putsp@ce{\number-\pslly} translate
}%
}}
\def\putsp@ce#1{#1 }
\def\psfordvitps{
\def\psdimt@n@sp##1{\d@mx=##1\relax\edef\psn@sp{\number\d@mx}}
\def\PSspeci@l##1##2{%
\special{dvitps: Include0 "psfig.psr"}
\psdimt@n@sp{\drawingwd}
\special{dvitps: Literal "\psn@sp\space"}
\psdimt@n@sp{\drawinght}
\special{dvitps: Literal "\psn@sp\space"}
\psdimt@n@sp{\psllx bp}
\special{dvitps: Literal "\psn@sp\space"}
\psdimt@n@sp{\pslly bp}
\special{dvitps: Literal "\psn@sp\space"}
\psdimt@n@sp{\psurx bp}
\special{dvitps: Literal "\psn@sp\space"}
\psdimt@n@sp{\psury bp}
\special{dvitps: Literal "\psn@sp\space startTexFig\space"}
\special{dvitps: Include1 "##1"}
\special{dvitps: Literal "endTexFig\space"}
}}
\def\psforDVIALW{
\def\PSspeci@l##1##2{
\special{language "PS"
literal "##2 1000 div dup scale"
include "##1"}}}
\def\psonlyboxes{
\def\PSspeci@l##1##2{%
\at(0cm;0cm){\boxit{\vbox to\drawinght
  {\vss
  \hbox to\drawingwd{\at(0cm;0cm){\hbox{(##1)}}\hss}
  }}}
}%
}
\def\psloc@lerr#1{%
\let\savedPSspeci@l=\PSspeci@l%
\def\PSspeci@l##1##2{%
\at(0cm;0cm){\boxit{\vbox to\drawinght
  {\vss
  \hbox to\drawingwd{\at(0cm;0cm){\hbox{(##1) #1}}\hss}
  }}}
\let\PSspeci@l=\savedPSspeci@l
}%
}
\def\psfordvi2ps{
\def\PSspeci@l##1##2{%
\special{insert size \number\drawingwd\space \number\drawinght\space 1sp ##1}%
}}
%
%
\newread\pst@mpin
\newdimen\drawinght\newdimen\drawingwd
\newdimen\psxoffset\newdimen\psyoffset
\newbox\drawingBox
\newif\ifNotB@undingBox
\newhelp\PShelp{Proceed: you'll have a 5cm square blank box instead of
your graphics (Jean Orloff).}
\def\@mpty{}
\def\s@tsize#1 #2 #3 #4\@ndsize{
  \def\psllx{#1}\def\pslly{#2}%
  \def\psurx{#3}\def\psury{#4}
  \ifx\psurx\@mpty\NotB@undingBoxtrue
  \else
    \drawinght=#4bp\advance\drawinght by-#2bp
    \drawingwd=#3bp\advance\drawingwd by-#1bp
    \advance\psyoffset by \drawinght
  \fi
  }
\def\sc@nline#1:#2\@ndline{\edef\p@rameter{#1}\edef\v@lue{#2}}
\def\g@bblefirstblank#1#2:{\ifx#1 \else#1\fi#2}
\def\psm@keother#1{\catcode`#112\relax}
\def\execute#1{#1}
{\catcode`\%=12
\xdef\B@undingBox{
}   
\def\ReadPSize#1{
 \edef\PSfilename{#1}
 \openin\pst@mpin=#1\relax
 \ifeof\pst@mpin \errhelp=\PShelp
   \errmessage{I haven't found your postscript file (\PSfilename)}
   \psloc@lerr{was not found}
   \s@tsize 0 0 142 142\@ndsize
   \closein\pst@mpin
 \else
   \immediate\write\psbj@inaux{#1,}
   \loop
     \executeinspecs{\catcode`\ =10\global\read\pst@mpin to\n@xtline}
     \ifeof\pst@mpin
       \errhelp=\PShelp
       \errmessage{(\PSfilename) is not an Encapsulated PostScript File:
           I could not find any \B@undingBox: line.}
       \edef\v@lue{0 0 142 142:}
       \psloc@lerr{is not an EPSFile}
       \NotB@undingBoxfalse
     \else
       \expandafter\sc@nline\n@xtline:\@ndline
       \ifx\p@rameter\B@undingBox\NotB@undingBoxfalse
         \edef\t@mp{%
           \expandafter\g@bblefirstblank\v@lue\space\space\space}
         \expandafter\s@tsize\t@mp\@ndsize
       \else\NotB@undingBoxtrue
       \fi
     \fi
   \ifNotB@undingBox\repeat
   \closein\pst@mpin
 \fi
\message{#1}
}
%
%
\newcount\xscale \newcount\yscale \newdimen\pscm\pscm=1cm
\newdimen\d@mx \newdimen\d@my
\let\ps@nnotation=\relax
\def\psboxto(#1;#2)#3{\vbox{
   \ReadPSize{#3}
   \divide\drawingwd by 1000
   \divide\drawinght by 1000
   \d@mx=#1
   \ifdim\d@mx=0pt\xscale=1000
         \else \xscale=\d@mx \divide \xscale by \drawingwd\fi
   \d@my=#2
   \ifdim\d@my=0pt\yscale=1000
         \else \yscale=\d@my \divide \yscale by \drawinght\fi
   \ifnum\yscale=1000
         \else\ifnum\xscale=1000\xscale=\yscale
                    \else\ifnum\yscale<\xscale\xscale=\yscale\fi
              \fi
   \fi
   \divide \psxoffset by 1000\multiply\psxoffset by \xscale
   \divide \psyoffset by 1000\multiply\psyoffset by \xscale
   \global\divide\pscm by 1000
   \global\multiply\pscm by\xscale
   \multiply\drawingwd by\xscale \multiply\drawinght by\xscale
   \ifdim\d@mx=0pt\d@mx=\drawingwd\fi
   \ifdim\d@my=0pt\d@my=\drawinght\fi
   \message{scaled \the\xscale}
 \hbox to\d@mx{\hss\vbox to\d@my{\vss
   \global\setbox\drawingBox=\hbox to 0pt{\kern\psxoffset\vbox to 0pt{
      \kern-\psyoffset
      \PSspeci@l{\PSfilename}{\the\xscale}
      \vss}\hss\ps@nnotation}
   \global\ht\drawingBox=\the\drawinght
   \global\wd\drawingBox=\the\drawingwd
   \baselineskip=0pt
   \copy\drawingBox
 \vss}\hss}
  \global\psxoffset=0pt
  \global\psyoffset=0pt
  \global\pscm=1cm
  \global\drawingwd=\drawingwd
  \global\drawinght=\drawinght
}}
%
%
\def\psboxscaled#1#2{\vbox{
  \ReadPSize{#2}
  \xscale=#1
  \message{scaled \the\xscale}
  \divide\drawingwd by 1000\multiply\drawingwd by\xscale
  \divide\drawinght by 1000\multiply\drawinght by\xscale
  \divide \psxoffset by 1000\multiply\psxoffset by \xscale
  \divide \psyoffset by 1000\multiply\psyoffset by \xscale
  \global\divide\pscm by 1000
  \global\multiply\pscm by\xscale
  \global\setbox\drawingBox=\hbox to 0pt{\kern\psxoffset\vbox to 0pt{
     \kern-\psyoffset
     \PSspeci@l{\PSfilename}{\the\xscale}
     \vss}\hss\ps@nnotation}
  \global\ht\drawingBox=\the\drawinght
  \global\wd\drawingBox=\the\drawingwd
  \baselineskip=0pt
  \copy\drawingBox
  \global\psxoffset=0pt
  \global\psyoffset=0pt
  \global\pscm=1cm
  \global\drawingwd=\drawingwd
  \global\drawinght=\drawinght
}}
%
\def\psbox#1{\psboxscaled{1000}{#1}}
%
%
%
\newif\ifn@teof\n@teoftrue
\newif\ifc@ntrolline
\newif\ifmatch
\newread\j@insplitin
\newwrite\j@insplitout
\newwrite\psbj@inaux
\immediate\openout\psbj@inaux=psbjoin.aux
\immediate\write\psbj@inaux{\string\joinfiles}
\immediate\write\psbj@inaux{\jobname,}
%
%
\immediate\let\oldinput=\input
\def\input#1 {
 \immediate\write\psbj@inaux{#1,}
 \oldinput #1 }
\def\empty{}
\def\setmatchif#1\contains#2{
  \def\match##1#2##2\endmatch{
    \def\tmp{##2}
    \ifx\empty\tmp
      \matchfalse
    \else
      \matchtrue
    \fi}
  \match#1#2\endmatch}
\def\warnopenout#1#2{
 \setmatchif{TrashMe,psbjoin.aux,psbjoin.all}\contains{#2}
 \ifmatch
 \else
   \immediate\openin\pst@mpin=#2
   \ifeof\pst@mpin
     \else
     \errhelp{If the content of this file is so precious to you, abort (ie
press x or e) and rename it before retrying.}
     \errmessage{I'm just about to replace your file named #2}
   \fi
   \immediate\closein\pst@mpin
 \fi
 \message{#2}
 \immediate\openout#1=#2}
{
\catcode`\%=12
\gdef\splitfile#1 {
 \immediate\openin\j@insplitin=#1
 \message{Splitting file #1 into:}
 \warnopenout\j@insplitout{TrashMe}
 \loop
   \ifeof
     \j@insplitin\immediate\closein\j@insplitin\n@teoffalse
   \else
     \n@teoftrue
     \executeinspecs{\global\read\j@insplitin to\spl@tinline\expandafter
       \ch@ckbeginnewfile\spl@tinline
     \ifc@ntrolline
     \else
       \toks0=\expandafter{\spl@tinline}
       \immediate\write\j@insplitout{\the\toks0}
     \fi
   \fi
 \ifn@teof\repeat
 \immediate\closeout\j@insplitout}
\gdef\ch@ckbeginnewfile#1
 \def\t@mp{#1}
 \ifx\empty\t@mp
   \def\t@mp{#3}
   \ifx\empty\t@mp
     \global\c@ntrollinefalse
   \else
     \immediate\closeout\j@insplitout
     \warnopenout\j@insplitout{#2}
     \global\c@ntrollinetrue
   \fi
 \else
   \global\c@ntrollinefalse
 \fi}
\gdef\joinfiles#1\into#2 {
 \message{Joining following files into}
 \warnopenout\j@insplitout{#2}
 \message{:}
 {
 \edef\w@##1{\immediate\write\j@insplitout{##1}}
 \w@{
 \w@{
 \w@{
 \w@{
 \w@{
 \w@{
 \w@{
 \w@{
 \w@{\string\input\space psbox.tex}
 \w@{\string\splitfile{\string\jobname}}
 }
 \tre@tfilelist#1, \endtre@t
 \immediate\closeout\j@insplitout}
\gdef\tre@tfilelist#1, #2\endtre@t{
 \def\t@mp{#1}
 \ifx\empty\t@mp
   \else
   \llj@in{#1}
   \tre@tfilelist#2, \endtre@t
 \fi}
\gdef\llj@in#1{
 \immediate\openin\j@insplitin=#1
 \ifeof\j@insplitin
   \errmessage{I couldn't find file #1.}
   \else
   \message{#1}
   \toks0={
   \immediate\write\j@insplitout{\the\toks0}
   \executeinspecs{\global\read\j@insplitin to\oldj@ininline}
   \loop
     \ifeof\j@insplitin\immediate\closein\j@insplitin\n@teoffalse
       \else\n@teoftrue
       \executeinspecs{\global\read\j@insplitin to\j@ininline}
       \toks0=\expandafter{\oldj@ininline}
       \let\oldj@ininline=\j@ininline
       \immediate\write\j@insplitout{\the\toks0}
     \fi
   \ifn@teof
   \repeat
   \immediate\closein\j@insplitin
 \fi}
}
\def\autojoin{
 \immediate\write\psbj@inaux{\string\into\space psbjoin.all}
 \immediate\closeout\psbj@inaux
 \input psbjoin.aux
}
%
%
%
%
\def\centinsert#1{\midinsert\line{\hss#1\hss}\endinsert}
\def\psannotate#1#2{\def\ps@nnotation{#2\global\let\ps@nnotation=\relax}#1}
\def\pscaption#1#2{\vbox{
   \setbox\drawingBox=#1
   \copy\drawingBox
   \vskip\baselineskip
   \vbox{\hsize=\wd\drawingBox\setbox0=\hbox{#2}
     \ifdim\wd0>\hsize
       \noindent\unhbox0\tolerance=5000
    \else\centerline{\box0}
    \fi
}}}
\def\psfig#1#2#3{\pscaption{\psannotate{#1}{#2}}{#3}}
\def\psfigurebox#1#2#3{\pscaption{\psannotate{\psbox{#1}}{#2}}{#3}}
%
\def\at(#1;#2)#3{\setbox0=\hbox{#3}\ht0=0pt\dp0=0pt
  \rlap{\kern#1\vbox to0pt{\kern-#2\box0\vss}}}
%
\newdimen\gridht \newdimen\gridwd
\def\gridfill(#1;#2){
  \setbox0=\hbox to 1\pscm
  {\vrule height1\pscm width.4pt\leaders\hrule\hfill}
  \gridht=#1
  \divide\gridht by \ht0
  \multiply\gridht by \ht0
  \gridwd=#2
  \divide\gridwd by \wd0
  \multiply\gridwd by \wd0
  \advance \gridwd by \wd0
  \vbox to \gridht{\leaders\hbox to\gridwd{\leaders\box0\hfill}\vfill}}
%
\def\fillinggrid{\at(0cm;0cm){\vbox{
  \gridfill(\drawinght;\drawingwd)}}}
%
%
\def\textleftof#1:{
  \setbox1=#1
  \setbox0=\vbox\bgroup
    \advance\hsize by -\wd1 \advance\hsize by -2em}
\def\textrightof#1:{
  \setbox0=#1
  \setbox1=\vbox\bgroup
    \advance\hsize by -\wd0 \advance\hsize by -2em}
\def\endtext{
  \egroup
  \hbox to \hsize{\valign{\vfil##\vfil\cr%
\box0\cr%
\noalign{\hss}\box1\cr}}}
%
\def\frameit#1#2#3{\hbox{\vrule width#1\vbox{
  \hrule height#1\vskip#2\hbox{\hskip#2\vbox{#3}\hskip#2}%
        \vskip#2\hrule height#1}\vrule width#1}}
\def\boxit#1{\frameit{0.4pt}{0pt}{#1}}
\catcode`\@=12 
%
\psfordvips   

\section{Introduction}

\label{Intro}

During the past few years, research carried out on chaotic eigenfunctions 
has provided very important results. Berry and Voros \cite{berry1, 
voros1} conjectured that, in the semiclassical limit, these eigenfunctions 
would be locally like random superpositions 
of plane waves; this conjecture is supported by theorems of Shnirelman 
\cite{shn} and Colin de Verdi\`ere \cite{colin}. But Heller \cite{heller1} 
found that a large number of highly excited eigenfunctions of the Bunimovich 
stadium billiard \cite{buni} have density enhancements along the shortest 
periodic orbits ({\bf PO}s). Since then several studies have focused on these
phenomena and led to theoretical developments \cite{bogoberry} and 
experimental observations such as in macroscopic billiard-shaped microwave 
cavities \cite{Kudro}, tunnel junctions \cite{fromwilk} and Hydrogen atoms in 
strong magnetic fields \cite{wintgen,main}. 

Recently, a variety of new 
aproaches to study the structure of chaotic eigenfunctions have been developed 
\cite{pol,kap,tom,ver1}. One of them consists of the semiclassical construction of
resonances with hyperbolic structure associated to unstable periodic 
orbits \cite{ver1}. These resonances were studied both analytically and 
numerically. They are classically motivated constructions that take into 
account complete classical information in the neighbourhood of a {\bf PO}. 
The so-called scar functions, as they have been named, can be obtained 
by a linear combination of resonances \cite{ver2,ver3} of a 
periodic orbit at a given energy with minimum energy dispersion. 
Resonances span a basis and are 
obtained by applying a creation operator over a vacuum state. The vacuum 
state (i.e. a resonance with no transverse excitations) is constructed with 
a (conveniently selected) Gaussian wave packet  
that follows a modified transverse motion along the chosen orbit. The modified 
motion, which is the result of dropping the pure hyperbolic one, describes a 
bunch of {\bf PO}s surrounding the chosen trajectory. For instance, the 
eigenvectors of the monodromy 
matrix evolve without the exponential contraction-dilation that is expected in 
this kind of hyperbolic dynamics. Instead, after one period they return to 
themselves (up to a minus sign in some cases). In the same way, 
the wave packet returns to itself with an accumulated phase 
that is an integral multiple of $2 \pi$, guaranteeing its continuity. 

Billiards are among the most interesting and well-studied systems in quantum 
chaos. We are going to work with the Bunimovich stadium billiard. For 
billiards, the boundary is the natural Poincar\'e section. We present a 
formulation of resonances and scar functions over the boundary. This reduction 
makes the calculations easier and it is a clear advantage when exploring a great 
number of eigenfunctions. 
Of course this is not the only reason to obtain these  
scar functions expressions, since they are more than tools for investigating 
numerically the structure of chaotic eigenfunctions. 
They are the cornerstone of the semiclassical 
theory of short periodic orbits of References \cite{ver2,ver3}. In 
this context it is possible to obtain all quantum information of a 
chaotic Hamiltonian system just by knowing classical information of a 
small number of short {\bf PO}s (in fact, the shortest ones, whose number 
increases at most linearly with a Heisemberg time). This is accomplished by 
evaluating the interaction between {\bf PO}s. In previous calculations \cite{ver3} 
only vacuum states of a handful of the shortest {\bf PO}s were used. 
For higher energies it is necessary to incorporate 
excitations to these vacuum states and also longer {\bf PO}s. All this can be 
done by using scar 
functions associated to a greater number of {\bf PO}s than the ones needed at 
low energy values. It is evident that working on the billiard domain will 
not be the best choice in these cases, and expressions on the boundary 
become essential. Hence, we need to develop an efficient method to evaluate them, 
this being one of the main goals of this article.  

We also go further by taking into account the non linear behaviour of the 
unstable and stable manifolds. Then, we are able to extend the validity 
of scar functions beyond the linear regime. The chosen approach is general 
in nature, and despite it is applied to the special case of the Bunimovich 
stadium it can be extended to general systems in a straight forward 
way. These new ingredients do not significantly alter the construction 
of scar functions, which preserve their compact character. These 
new expressions provide a more comprehensive description of the phase space 
in the neighbourhood of each orbit. They are the other main goal 
of the present work. 

This paper is organised as follows, Section 2 consists of a 
detailed explanation of the construction of resonances on the billiard 
domain. Here we address many 
subtleties involved in the wave function calculations and local expressions 
are provided.  
Moreover, a comprehensive approach is offered for resonances including vacuum 
states and excitations in a single formulation. 
Though the construction may seem complex the great advantage is that it 
is a general one, suitable for any billiard. In Section 3 the construction is 
extended with expressions on the boundary. In this part explicit formulas 
amenable to extensive and high energies 
calculations are presented. Also a brief summary of the results from 
the previous section and their application to the construction of scar 
functions on the billiard boundary is developed; here, we give explicit 
examples. Finally, Section 4 is devoted to conclusions.    						
			
\section{Resonances on the billiard domain}

\label{domain}

In this section we give a thorough explanation of 
the construction of the resonances associated to a given 
trajectory $\gamma$ of length $L$ belonging to the 
Bunimovich stadium. We define a coordinate $x$ along the 
trajectory and a coordinate $y$ perpendicular to it (such that 
$y=0$ defines the orbit). The evolution of an initial displacement 
$(y, p_y)$ at $x \, = \, 0$ into 
$(y(x), p_y(x))$ at $x$ can be obtained by means of the symplectic 
stability matrix $M(x)$ with elements $m_{ij}$, 
\[
\left( \begin{array}{c} 
		y(x) \\
		p_{y}(x)
	         \end{array} \right)
= \left( \begin{array}{cc}
	    m_{11}(x)   &   m_{12}(x)  \\
	    m_{21}(x)   &   m_{22}(x)
		\end{array}  \right)
\left( \begin{array}{c}
	       y   \\
	       p_{y}
		\end{array}   \right).
\]
The eigenvectors $\xi_u$ and $\xi_s$ of $M$ give the unstable and stable directions 
associated to the orbit. These vectors evolve as $\;\tilde{\xi}_u(x)=M(x)\;\xi_u\;$. 
After one period they return over themselves, 
$\tilde{\xi}_u(L)=(-1)^{\mu}\; e^{\lambda L} \;\xi_u$ and
$\tilde{\xi}_s(L)=(-1)^{\mu}\; e^{-\lambda L} \;\xi_s$, where $\mu$ is the total 
number of half turns made by $\tilde \xi_u(x)$ and $\tilde \xi_s(x)$ during 
its evolution along the orbit. For a billiard this 
number corresponds 
to $\mu = \nu + N_{ref}$ where $\nu$ is the maximum number of conjugated points 
(in particular, for the stadium $\nu$ is exactly the number of bounces with the circle). 
$N_{ref}$ is the total number of reflections with the billiard boundary and 
the symmetry lines. This point will be addressed later when we refer 
to quantisation conditions. 

We are going to decompose the motion given by $M(x)$ into a purely hyperbolic 
and a periodic one. In order to do so we need to specify the contraction-dilation 
rate along the manifolds. This can be done by first finding one of the $x_0$ points 
on the trajectory where the projections of $\xi_u(x_0)$ and $\xi_s(x_0)$ on 
$y$ and $p_y$ are equal in absolute value, i.e., the unstable and stable 
directions would be symmetrical with respect to the axes. There are  
$2 \nu$ points like these on the orbit. They can be found by using the 
fact that $M_{x_0}$ (the return map starting at $x=x_0$) has equal diagonal 
elements when this condition on the eigenvectors is met. By means of the
relation $M_{x_0} = M(x_0) M(L) M(x_0)^{-1}$ this condition can be easily implemented.

Then, we decompose $M(x)$ into a periodic matrix $F(x)$ describing the evolution 
of the manifolds and a matrix which is responsible of the contraction-dilation 
along them. This is just an application of Floquet theorem \cite{Yaku}: 
\begin{equation}
M(x)=F(x)\; \exp [f(x)\; \lambda \;K],
\label{de}
\end{equation}
where $K \equiv BDB^{-1}$ with $D$ a diagonal matrix of elements 
$d_{11}=1$ and $d_{22}=-1$ and $B=(\xi_{u}\;\xi_{s})$, i.e., the symplectic 
matrix transforming coordinates from the new axes $\xi_{u}$ and $\xi_{s}$ 
to the old ones $y$ and $p_y$. 
The real function $f(x)$ (required to fulfill $f(0)=0$ and $f(L)=L$) can be 
seen as the relation between the lengths of $\tilde \xi_u(x)$ and 
$\xi_u$, but the plane $y$-$p_y$ has no defined norm so establishing this 
relation is in general 
impossible. Further 
conditions can be imposed on $f(x)$, (see Ref. \cite{ver1}). 
However, without loss of generality, 
we are going to consider the easier choice 
$f(x)=x$ for the present calculations.
On the other hand the neutral motion given by $F(x)$ can be obtained by its 
action on $\xi_u$ and $\xi_s$, allowing us to define the set 
$y_u(x)$, $y_s(x)$, $p_u(x)$ and $p_s(x)$ of periodic functions: 
\begin{equation}
\left( \begin{array}{c}
	 \!\!  y_{u}(x) \!\! \\
	 \!\! p_{u}(x) \!\!
	\end{array} \right) \equiv \xi_{u}(x)
\equiv F(x)\;\xi_{u}=e^{-f(x)\lambda}\; M(x)\;\xi_{u},
\label{def1}
\end{equation}
and
\begin{equation}
\left( \begin{array}{c}
\!\!  y_{s}(x) \!\! \\
\!\! p_{s}(x) \!\!
\end{array} \right)\equiv \xi_{s}(x)\equiv 
F(x)\;\xi_{s}=e^{f(x)\lambda}\; M(x)\;\xi_{s}.
\label{def2}
\end{equation}

Since $F(x)$ is area preserving, these functions satisfy 
the sympletic property $ y_{u}(x)p_{s}(x)-
y_{s}(x)p_{u}(x)=\xi_{u}(x) \wedge \xi_{s}(x)= 
\xi_{u} \wedge \xi_{s}=J$ ($J$ being the unit of action in the $y$-$p_y$ 
plane).

A resonance can be essentially seen 
as the product of a plane wave in the $x$ direction, being the 
semiclassical approximation for the unidimensional motion 
along the orbit, and a Gaussian wave packet in the transverse 
coordinate, which follows a dynamics without dilation-contraction 
along the unstable and stable manifolds of the trajectory.  
The vacuum state (semiclassically normalised to unity) is given by \cite{ver1}: 
\begin{equation}
\psi_{\gamma}^{(0)}(x,y)\!=\!\frac{\exp \{i\;[S(x)+y^{2}\;\Gamma(x)/2]/
\hbar-i \;\phi(x)/2\} } {\sqrt{T \;\dot x}\;\;\; 
[\pi\;(\hbar/J)\; 
|Q(x)|^{2}]^{1/4}},
\label{res}
\end{equation}
where $T$ is 
the period of the orbit and $\phi(x)$ is the angle swept by $Q(x)$ while 
evolving from $0$ to $x$.  In this expression $\Gamma(x) = P(x)/Q(x)$, 
where $Q(x)$ and $P(x)$ are
the components of a complex vector constructed with the stable 
and unstable manifolds. These components are obtained as:
\begin{equation} 
\left( \begin{array}{c}
		    \!\!  Q(x)=y_{u}(x)+\!i\; y_{s}(x) \!\! \\
		  \!\!  P(x)=p_{u}(x)+\!i\; p_{s}(x) \!\!
			\end{array} \right)
\equiv\xi_{u}(x)+i \;\xi_{s}(x)  =M(x)B \left( \begin{array}{c}
		    \!\!  e^{-f(x) \lambda} \!\! \\
		  \!\! i \; e^{ f(x) \lambda} \!\!
			\end{array} \right). 
\label{vec}
\end{equation}

For billiards we take a slightly modified evolution matrix ($\tilde M(x)$) 
in order to have a continuous $Q(x)$; since the phase $\phi(x)$ of $Q(x)$ 
must be known in detail for resonances construction, continuity is a very 
reasonable condition to ask for. This can be done by means of 
$\tilde M(x) \equiv (-1)^{N(x)} M(x)$, with $N(x)$ being the number 
of reflections while evolving from $0$ to $x$. In turn, this matrix can 
be constructed by using two types of matrices: 
\[
M_{1}(l)=\left( \begin{array}{cc}
	    1   &   l  \\
	    0   &   1
		\end{array}  \right),\;\;{\rm and}\;\;\;
M_{2}(\theta)=\left( \begin{array}{cc}
	    1   &   0  \\
	    -2/ \cos(\theta)   &   1
		\end{array}  \right).
\]
$M_1(l)$ describes the evolution for a path of length $l$ without bounces 
with the circle (the transverse momentum is measured in units of the 
momentum along the trajectory). $M_2(\theta)$ takes into account a bounce 
with the circle ($\theta$ defines the angle between the incoming trajectory 
and the radial direction). In the following we assume this matrix instead 
of the original one given for the most general expressions in the previous 
theoretical introduction. The reflections related phase will be included 
in the expressions directly.

Taking the wave functions defined in Eq (\ref{res}) as the vacuum state 
for appropriate creation-anihilation operators (see \cite{ver1}), the following 
expression results for a resonance with $m$ transverse excitations:

\begin{equation}
\psi_{\gamma}^{(m)}(x,y)=\frac{e^{-im\phi(x)}}{\sqrt{2^{m}\;m!}}\;
H_{m}\!\left[ \frac{y\sqrt{J/\hbar}}{|Q(x)|} \right]\; 
\psi_{\gamma}^{(0)}(x,y),
\label{excita}
\end{equation}
where $\phi(x)$ is the phase introduced in Eq (\ref{res}), and
$H_{m}(z)$ are the Hermite polynomials 
($H_0 = 1$, $H_1 = 2 \xi$, $H_2 = 4 \xi^2 - 2$, $\ldots$). 
It is easy to see that $\psi_{\gamma}^{(m)}$ is also a product of two
functions; the solution for the motion along the orbit 
and $m$ excitations of a transverse
Gaussian wave packet which evolves following the same periodic 
motion previously mentioned. 

Inside each family, resonances are identified by the integer number 
$n=0,1,\ldots$, the number of excitations along the orbit 
and by $m=0,1,\ldots$, the transverse excitations. The wave number 
$k$ depends on $\gamma$, $n$ and $m$ through the rule: 
\begin{equation}
S(L)/\hbar-N_b \pi/2-(m+1/2) \mu \pi = 2 \pi n,
\label{bohr1}
\end{equation}
which guarantees the continuity of $\psi_\gamma^{(m)}$ at $x=L=0$.  
In this expression $S(L)=\int_0^L p_x dx$ is the dynamical action and 
$\mu$ is the topological phase; that is, the number of half turns made 
by the manifolds along the orbit. Finally, $N_b$ is a pure quantum phase 
related to the boundary conditions (see \cite{ver1}), which 
is equal to the number of reflections 
satisfying Dirichlet boundary conditions minus the number of reflections 
satisfying Neumann conditions.  

Rule (\ref{bohr1}) can be divided into the $m \mu$ 
even and odd cases: 
\begin{equation}
S(L)/\hbar-N_{b} \pi/2  -\mu \pi/2= 2 \pi\; (n+m \mu/2)
\label{bohr2}
\end{equation}
for $m\mu$ even, and
\begin{equation}
S(L)/\hbar-N_{b} \pi/2  -3\mu \pi/2= 2 \pi\; [n+(m\!-\!1) \mu/2],
\label{bohr3}
\end{equation}
for the odd case. For the desymmetrized stadium billiard it results in 
\[
S(L)/\hbar \; = \; L k \, \, {\rm ;} \, \, N_b \; = \; N_s+s_h N_h + s_v N_v 
\]
and $\mu = N_{ref} + \nu$ with $N_{ref}=N_s+N_h+N_v$ where $N_s$, $N_h$ 
and $N_v$ are the number of reflections with the stadium boundary and 
the horizontal and vertical symmetry lines respectively. Finally, 
$\nu$ is the number of bounces with the quarter of the circle. 
The value of $s_{h}$ is $0$ or $1$ depending on the symmetry with 
respect to the horizontal axis being even or odd; 
this is equivalent for $s_{v}$, where the vertical symmetry is 
considered. The set of allowed $m$ values can range from 
$0$ or $1$ (depending on $m \mu$ being even or odd) to $m<A_{eff}/2 \pi 
\hbar$ (where $A_{eff}$ is the transverse area in which the construction 
is valid). For simplicity, we are going to 
consider $0(1) \leq m < n$ as a satisfactory criterion.

The time reversal properties of the system under study can be applied 
to Eq (\ref{excita}), giving a real resonance expression.
Resonances are constructed explicitely for the stadium billiard by 
assigning a semiclassical expression to each 
straight line of the orbit. The first one of these lines is the segment 
of $\gamma$ that begins in $x_{1}=0$. Let $x_{2}$ ($>x_{1}$) be the value 
of $x$ where the path reaches the border of the stadium after the 
corresponding evolution. The path departing from $x_{2}$ defines the 
second line and so on until $x=L/2$ is reached. 
If we define local coordinates $(x^{(j)},y^{(j)})$ over each line in 
such a way that $x^{(j)}=x$ inside the desymmetrized billiard, the 
expression for the $j$th line turns out to be [we are going to use 
$(x,y)$ in the following expressions, understanding that they represent 
the variables $(x^{(j)},y^{(j)})$ of the $j$th line]
\begin{eqnarray}
\psi_{j}^{(m)} (x,y) & = & \left( \frac{kR}{|Q(x)|^2} \right)^{1/4} \; 
f^{(m)} \; \left[ \frac{\sqrt{kR} \, y}{|Q(x)|} \right] 
 \nonumber \\
  &   &  \frac{2}{\sqrt{L}} \sin [k y^{2} g_{j}(x)+kx- N_b(x_j^{+}) 
  \pi/2 - \nonumber \\
  &   &  - (m+1/2) \Phi_{j}(x)-F_0-m G_0].
\label{realres}
\end{eqnarray}
Here $N_b(x_j^{+})$ is defined in the same way as $N_b$ but taking 
account of the bounces up to the point $x_j$ (including it), the 
corresponding term takes into account the quantum phase associated to 
boundary conditions and this is the resonance counterpart of the phase 
considered in the quantisation conditions.
Furthermore, 
\begin{equation}
g_{j}(x) \equiv Re [\Gamma(x)/2]= 
\frac{y_u \; p_u + y_s \; p_s}{2 \; (y_u^2+y_s^2)} \; , 
\label{geqn}
\end{equation}
where the $x$ dependence is understood. 
We would like to point out that the factor $k$ in the term 
$k y^2 g_j(x)$ in Eq (\ref{realres}) is different from the 
factor $1/\hbar$ in Eq (\ref{res}). This is due to the fact that  
momenta in the expression for $g_j(x)$ are measured in units 
of $\hbar k$, i.e., the momentum along the trajectory.
The oscillator functions are defined by
\[
f^{(m)} (\xi)= \pi^{-1/4} (2^m m!)^{-1/2} 
e^{-\xi^2/2} H_m(\xi), 
\]
where $H_m$ is the $m$th degree Hermite 
polynomial. 
The $sine$ function in Eq (\ref{realres}) was 
chosen in order to obtain real resonances by virtue of the time 
reversal symmetry of the system. 
We have taken $T \dot x=L$ in Eq (\ref{res}) and considered 
$J=\hbar kR$ in Eq (\ref{excita}), with $k$ the wave number from 
the quantisation conditions and $R$ the circle radius of the stadium. 
Finally   $\Phi_{j}(x)= N_{ref}(x_{j}^{+}) \pi 
+[\tilde \varphi_{j}+\phi_{j}(x)+\alpha_{j}(x)]$, 
with 
$ \tilde \varphi_{j} = \tilde \varphi_{j-1} + \phi_{j-1}(x_j) + 
\alpha_{j-1}(x_j)$ for $j \geq 2$ and $ \tilde \varphi_{1} = 0$, follows 
the actual phase of $Q(x)$. 
In the first term of this expression we introduce the jumps in the 
sign of $Q(x)$ due to reflections on hard walls, that were avoided 
by using the $\tilde M(x)$ matrix instead of the right one. In fact, 
this is the only place where this change is needed because $\Gamma$ keeps 
its sign. The remaining term of this function, i.e., 
$\tilde \varphi_{j}+\phi_{j}(x)+\alpha_{j}(x)$ is explained in the 
subsection \ref{phasefollow}.

The initial phase of the resonances $\psi_{j}^{(m)}$, Eq (\ref{realres}), 
is essential to satisfy the boundary conditions. We 
have set $\tilde \varphi_1=0$ in the relations that define 
$ \tilde \varphi_{j}$, and account for the initial phase by means of 
two quantities, $F_0$ and $G_0$, which take values as shown in 
Fig. \ref{initialphase}. $F_0$ accounts 
for the initial phase in order to obtain an even (odd) function when 
symmetry is even (odd), without considering the symmetry of 
the oscillator functions (this will be addressed by $G_0$). 
In the case of having only one path to 
consider at the initial bounce, like in the first diagrams 
for both rows, the corresponding function must obey the symmetry 
by itself. In the second case of the first 
row  the symmetry related path is added but multiplied by $-1$ (see 
the final sum in this section and explanation thereby). 
Finally, in the two remaining cases (of the second row) 
an even function is needed since the function obtained by adding 
the symmetry related path multiplied by $-1$ ($1$) is 
(anti)symmetric. $G_0$ takes into account the symmetry of the 
Hermite polynomials, which goes like $m$, the number of transverse 
excitations. Its value depend on the starting angle 
of the trajectory. If this angle is zero, the symmetry of the resonance is 
the one given by the $F_0$ choice. In the other case, the symmetry 
of the oscillator function (i.e. Hermite polynomials) comes into play 
and in order to have the right final result we need to consider it. 
This can be easily accomplished by the choice of the $G_0$ value 
according to Fig. \ref{initialphase}. 

\begin{figure}[h]
\psyoffset=-20cm
\psxoffset=15cm
\psboxto(5cm;5cm){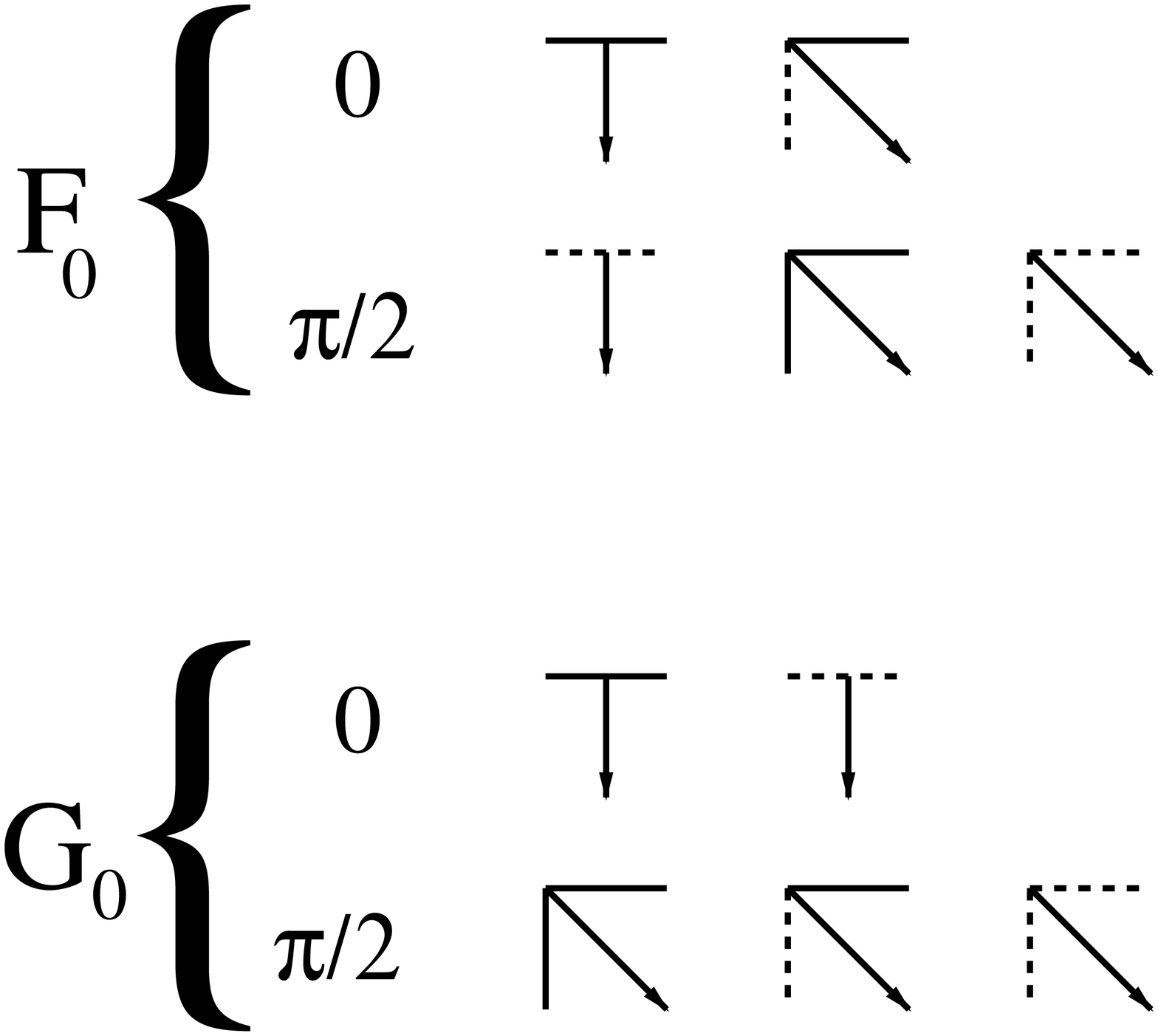}
\caption[]{\label{initialphase}$F_0$ and $G_0$ values depending on the 
type of starting point and on the boundary conditions. Dashed (solid) lines 
correspond to the symmetrical (antisymmetrical) condition, i.e., 
$s_h$, $s_v = 0 (1)$. Finally, arrows stand for the trajectories.}
\end{figure}

The transformation of local coordinates $(x^{(j)},y^{(j)})$ over the 
$j$ line to $(X,Y)$ coordinates that belong to the horizontal and 
vertical 
directions respectively, can be obtained in a simple way by means of an 
usual transformation. If $(X_{j},Y_{j})$ are the coordinates of the 
point 
$x_{j}$, and $\alpha_{j}$ is the angle between the $j$ line and the 
horizontal direction, $(x^{(j)}-x_j,y^{(j)})=G_{j}(X,Y)$ is given by 
\[
G_{j}(X,Y)=(X-X_{j},Y-Y_{j}) 
\left( \begin{array}{cc}
           \cos (\alpha_{j})   &  -\sin( \alpha_{j})  \\
           \sin( \alpha_{j})   &  \cos (\alpha_{j})
                \end{array}  \right) .
\]

Finally, the family of resonances $\psi_{\gamma}$ is constructed by 
means 
of all the lines including symmetries (see Fig. \ref{construccion} 
where 
this procedure is illustrated for one of the shortest periodic orbits 
of the stadium billiard), this avoids unnecesary evaluation of reflection 
points over the symmetry lines, considering only those over the boundary: 

\begin{equation}
\psi_{\gamma}(X,Y)=\sum_{j} \sum_{i=1}^{m_{h}} \sum_{l=1}^{m_{v}}
h_{i}\; v_{l}\; \psi_{j}[(x_j,0)+G_{j}(s_{l}X,s_{i}Y)].
\label{ec2}
\end{equation}
where $s_{i}\equiv(-1)^{i+1}$ and $s_{l}\equiv(-1)^{l+1}$.
$h_{i}=[\delta_{i,1}+\delta_{i,2}(1-2s_{h})]$ and
$v_{l}=[\delta_{l,1}+\delta_{l,2}(1-2s_{v})]$. 
$m_{h}$ and $m_{v}$ depend on $j$ and are specified like this: 
$m_{h}=1$ or $2$ if the line is symmetrical or not with respect 
to the horizontal axis respectively, and the same happens with the 
number $m_{v}$ for the vertical axis, although $m_{h}=2$ 
and $m_{v}=1$ if the line goes through the origin. This is the only 
considered choice of the two possible ones (the other is 
$m_{h}=1$ and $m_{v}=2$) such that the right number of lines 
are included in the case of having symmetry with respect to the origin.

\begin{figure}[h]
\psyoffset=-8cm
\psxoffset=1cm
\psboxto(10cm;10cm){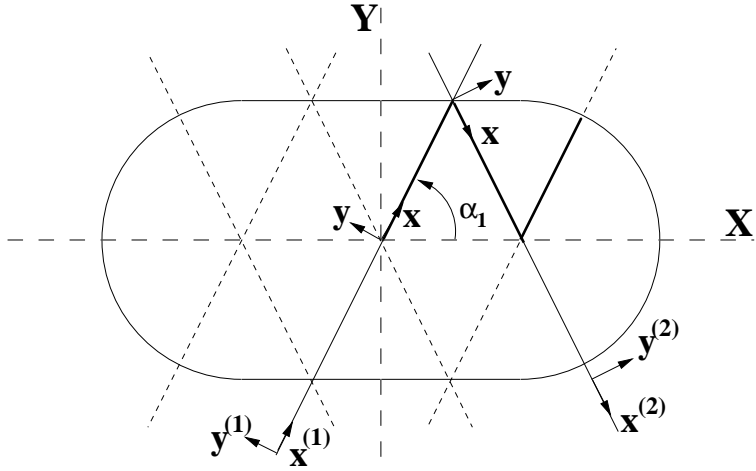}
\caption[]{\label{construccion} Set of lines including symmetries 
employed in the construction of resonances associated to 
one of the shortest periodic orbits. 
Different coordinate sets involved in this construction 
are shown.}
\end{figure}

This construction of resonances allows us to reduce 
calculations by exploiting the 
spatial symmetries of the system. We will explain the role of all 
these indices and coodinate rotations. The values of the function 
$\psi_{\gamma}$ in the general coordinates $(X,Y)$ in the domain are 
obtained from their corresponding ones in local coordinates over each 
line. This is the reason why in the case of lines having no symmetry 
with respect to the axis $X$ or $Y$, $m_h$ and $m_v$ adopt the value 
2 in such a way that the two symmetry related lines are included (in the 
case of showing that symmetry, $m_h$ and $m_v$ have value 1 and the 
only line considered is the original one). Then, the values given to 
$s_l$ and $s_i$ complete this process by reflecting the corresponding 
coordinate. 
The main point about this mechanism in Eq (\ref{ec2}) is that when second 
lines are involved the boundary 
conditions at the axes have to be met also. To achieve this result 
they are added with positive or negative sign if the 
symmetry is even ($s_h$ or $s_v$ equal to 0) or odd 
($s_h$ or $s_v$ equal to 1). This can be directly verified by 
replacing $i$ and $l$ by 2 in the formulas for 
$h_i$ and $v_l$, and also $s_h$ and $s_v$ by their corresponding values 
(always referring to the reflected lines). To satisfy the boundary 
conditions 
when a line goes through the origin of $(X,Y)$, only one reflection 
must 
be considered, this is the reason why this case is treated separately 
in our general rules of assignment for the $m_h$ and 
$m_v$ values.

Now that we have completed the construction of resonances on the domain of 
the billiard it is convenient to introduce the expression for the hyperbolic 
part of the Hamiltonian applied to these functions. As shown in Ref. 
\cite{ver1} it has a simple form in terms of conveniently defined 
creation-anihilation operators and also can be expressed as: 

\begin{eqnarray}
\frac{(\hat H - E)}{\hbar^2/2 {\rm M}} \; \psi_j^{(m)} \; & = & \;
- \alpha (x) \; [ \sqrt{(m+1)(m+2)} \;
\psi_j^{(m+2)} \; + \nonumber \\
 &  &  + \; \sqrt{(m-1) m} \; \psi_j^{(m-2)}],
\label{Hdom}
\end{eqnarray}
with $\alpha(x) = s \; 2{\rm M} \dot x / \hbar \; f'(x) \; \lambda /2 = 
s \; k \; f'(x) \; \lambda$ (where $f'(x)=1$, $s$ is the sign of the 
slope of the unstable direction with respect to the $y$ axis at 
the initial point $x=x_0$, and M the mass of the particle). 
The case $m=0$ corresponds to the 
expression given in \cite{ver2}. This formula or its version on the 
boundary which is derived in Section \ref{resboundary}, 
is the cornerstone of the short periodic orbits theory and 
also plays a crucial role in the construction of scar functions.

\subsection{Following the phase of $Q(x)$: a local expression for resonances}

\label{phasefollow}

We have pointed out that function $\Phi_{j}(x)$ is the 
phase accumulated by $Q(x)$ from the point $x_1=0$ to $x$ over the 
trajectory, 
followed in a continuous way and incorporating jumps due 
to rigid walls afterwards (we isolated them by using $\tilde M(x)$). 
In this subsection we are 
going to explain the calculation of the continuous part in detail. 

The function $\phi_{j}(x)={\rm arg}[Q_{j}(x)]-{\rm arg}[Q_{j}(x_{j})]$, 
where arg takes the argument of a complex number in the interval 
$[0;2 \pi)$, follows the phase of $Q(x)$ after the bounce in $x_j$. 
We must be careful in the analysis because the phase of $Q_j(x)$ 
is divided by 2. Then, the proposed method will be to consider the 
phase of $Q_j(x)$ in an arbitrary point of the line $j$. 
The complex vectors $Q(x)$ and $P(x)$ are obtained by
\[
\left( \begin{array}{c} 
                Q_{j}(x) \\
                P_{j}(x)
                 \end{array} \right)
= M_{1}(x-x_{j}) \left( \begin{array}{cc}
            a_{j}   &   b_{j}  \\
            c_{j}   &   d_{j}
                \end{array}  \right)
\left( \begin{array}{c}
               e^{-(x-x_{j})\lambda}   \\
              i\; e^{(x-x_{j})\lambda} 
                \end{array}   \right),
\]
where
\[
\left( \begin{array}{cc}
            a_{j}   &   b_{j}  \\
            c_{j}   &   d_{j}
                \end{array}  \right)
=\tilde{M}(x_{j}^{+}) B \left( \begin{array}{cc}
            e^{-x_{j}\lambda}   &   0  \\
            0   &   e^{x_{j}\lambda}
                \end{array}  \right)
= \left( \begin{array}{cc}
            y_{u}(x_j)   &   y_{s}(x_j)  \\
            p_{u}(x_j)   &   p_{s}(x_j)
                \end{array}  \right).
\]
Here $x_{j}^{+}$ means (for $j\!\geq \!2$) that $\tilde{M}$ (the stability 
matrix multiplied by $(-1)^{N_b(x_j^{+})}$) is evaluated 
after the bounce over the boundary in $x_{j}$.  
It can be easily verified that 
\begin{eqnarray}  Q_j(x) \; & = & \; [a_j+(x-x_j) c_j] 
\exp[- \lambda (x - x_j)] + \nonumber \\
 & & + \; i [b_j +(x-x_j) d_j] \exp[\lambda (x - x_j)].
\end{eqnarray}
The first conclusion that we can derive from this expression is that 
for $x \rightarrow -\infty$, $Q(x)$ is real, with sign equal to 
${\rm sgn}(-c_j)$, and for $x \rightarrow +\infty$ is pure imaginary, with 
its sign given by ${\rm sgn}(d_j)$. We will define the new variable 
$\tilde x=x -x_j$. With this 
definition 
we can easily write $Q_j(\tilde x)=(a_j+\tilde x c_j) \exp(- \lambda 
\tilde x) + i (b_j +\tilde x d_j) \exp(\lambda \tilde x)$. We are 
going to divide the $\tilde x$ axis, having in mind the 
places where the real and imaginary part of $Q(\tilde x)$ change 
signs. These are located at $\tilde x _{Re} = -a_j/c_j$ for the real 
part and at $\tilde x _{Im} = -b_j/d_j$ for the imaginary one.
It is easy to see that $Q(\tilde x _{Re})$ is pure imaginary with 
sign given by ${\rm sgn}[(b_j c_j-a_j d_j)/
c_j]={\rm sgn}(-1/c_j)={\rm sgn}(-c_j)$. 
On the other hand, $Q(\tilde x _{Im})$ is real and ${\rm sgn} \; 
Q(\tilde x _{Im})= {\rm sgn}[(d_j a_j-b_j c_j)/d_j]={\rm sgn}
(1/d_j)={\rm sgn} \; d_j$.
Moreover, once the signs of $c_j$ and $d_j$ are given, the order 
relation between $\tilde x _{Re}$ and $\tilde x _{Im}$ is specified. 
For instance, for $c_j>0$ and $d_j>0$, and using the fact that 
$a_j \; d_j - b_j \; c_j = 1$ there results 
$\tilde x _{Re} = -a_j/c_j < -b_j/d_j = \tilde x _{Im}$. 

The only idea that motivates 
the previous reasoning is the continuous evolution of $Q(\tilde x)$, 
without necessarily being a monotonous one. Hence, we can have four 
possible sign combinations and orderings of $\tilde x _{Re}$ and 
$\tilde x _{Im}$, as shown in Fig. \ref{phasesign}. 
The same Figure also shows the evolution of $Q(\tilde x)$ 
in the complex plane.

\begin{figure}[h]
\psyoffset=0cm
\psxoffset=1cm
\psboxto(10cm;10cm){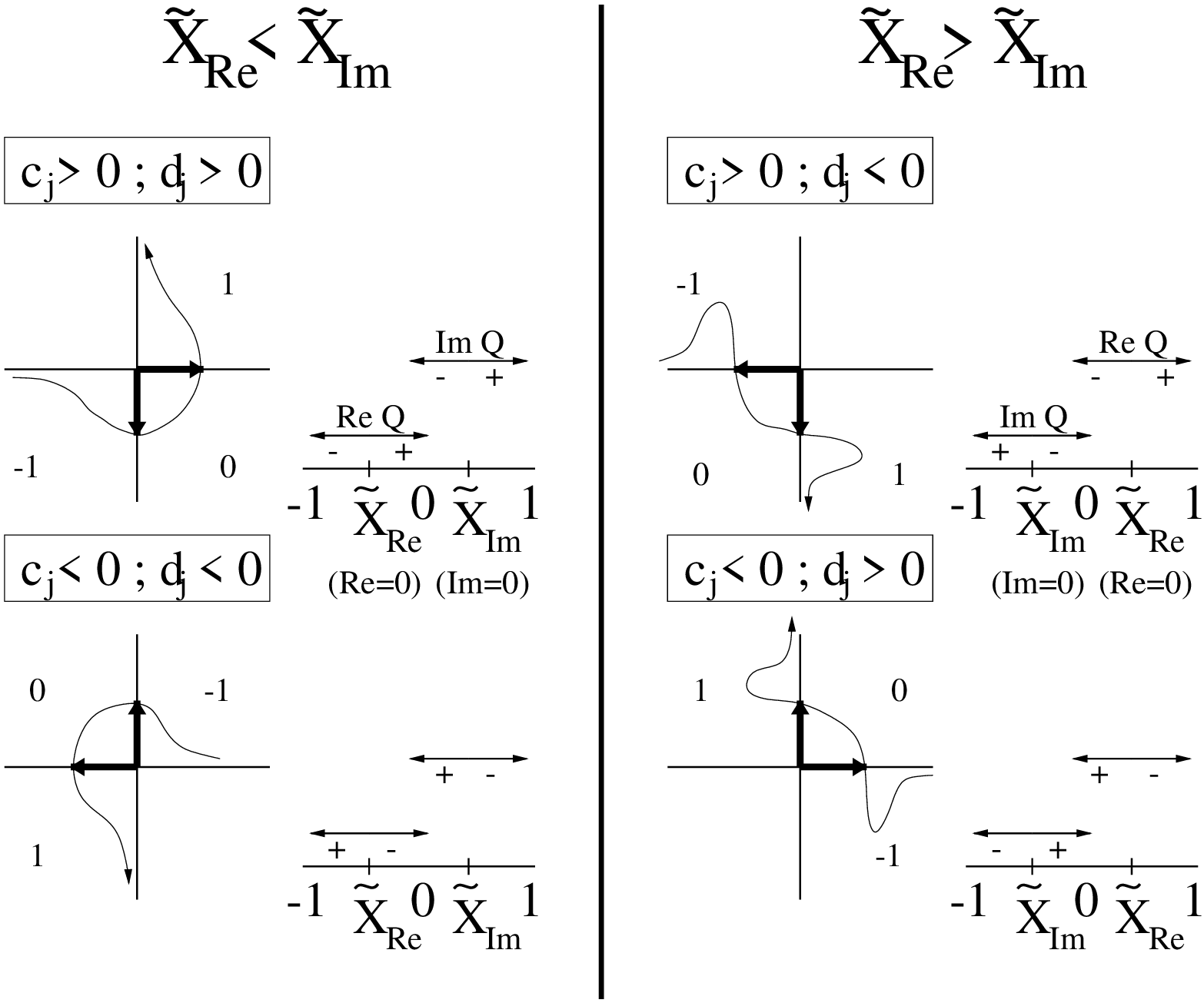}
\caption[]{\label{phasesign} Detail of the possible evolutions of the 
complex number $Q(\tilde x)$. The curves in three out of the four 
quadrants depict its time evolution. 
$\tilde x_{Re}$ divides the $\tilde x$ axis in domains with a positive or 
negative value of the real part of $Q(\tilde x)$. The same applies to $\tilde 
x_{Im}$ but for the imaginary part. Numbers $-1$, $0$ and $1$ denote the 
quadrants visited (in the given order) by $Q(\tilde x)$.}
\end{figure}

In these figures we have labelled the three 
sectors of the $\tilde x$ axis that are defined by 
$\tilde x_{Re}$ and $\tilde x_{Im}$ with $-1$, $0$ and $1$. 
They are the three quadrants where $Q(\tilde x)$ time evolution 
takes place, evolving from $-1$ to $0$ and then to $1$. 
This labelling can be defined through the function:

\[
n_{j}(\tilde x)=\left\{ \begin{array}{rl}
1 & {\rm if}\; \tilde x  >{\rm max}(-a_{j}/c_{j},-b_{j}/d_{j}) \\
-1 & {\rm if}\; \tilde x  <{\rm min}(-a_{j}/c_{j},-b_{j}/d_{j}) \\
0  & {\rm otherwise}.    
                                        \end{array} \right.
\]

Though rather technical, introducing $n_{j}$ allows us to define the 
resonances in a local form, and this is a very important feature of 
them. The curve representing  
$Q(\tilde x)$ (in Fig. \ref{phasesign}) is only qualitative and, 
of course, it does not imply monotony in the evolution. In the case 
of Figures corresponding to 
$\tilde x_{Re} > \tilde x_{Im}$, for instance when $d_j <0$, the 
imaginary component of $Q$ in the $-1$ region grows while $\tilde x$ 
decreases in absolute value; but in some point it must begin to 
decrease in absolute value in order to pass through zero in a continuous way, 
following the signs of the quadrant. 
Something similar happens in the other cases for both 
asymptotic limits. The arrows over the 
real and imaginary axes in these four schemes represent $Q(\tilde 
x_{Re})$ if they are over the imaginary axis and $Q(\tilde x_{Im})$ if they are 
over the real axis. 

Let´s put again the analysis in terms of the $x$ variable over the 
trajectory. Not all the changes of quadrant imply a jump in the phase of 
$Q(x)$ given by the arg function. While it is true that {\em inside} of 
each quadrant there is {\em no} need of monotony, it {\em is} needed among 
the quadrants (this is a kind of discrete monotony that allows us to say 
that the phase of $Q(x)$ is greater in 1 than in 0, and it is greater in 0 
than in -1). Let´s suppose now that there has been a change of 
region (or quadrant). In the case of $\tilde x >0$ (there has been an 
evolution from the point $x_j$) but $\phi_{j}(x)<0$ (the phase of 
$Q(x)$ relative to the one that it had in $x_j$ is lower), or 
$\tilde x <0$ and $\phi_{j}(x)>0$ (there has been a backward evolution 
but the phase grows), we are in a situation where the real positive 
axis has been crossed. Thus, we have to add or substract $2 \pi$, 
respectively, in order to continuously keep track of the phase 
(because if $\tilde x >0$ there has been an evolution from $x_j$ and 
the crossing has been in the counterclockwise or positive sense, 
and it is negative if $\tilde x <0$).
To summarize, if $n_{j}(x)\neq n_{j}(x_{j})$
and $(x-x_{j}) \phi_{j}(x)<0$ a phase $\alpha_{j}(x)=2\pi\; {\rm 
sgn}(x-x_{j})$ 
must be added; otherwise $\alpha_{j}(x)=0$.
If $c_{j}\!=\!0$ or $d_{j}\!=\!0$, $x_{j}$ can be replaced by any other 
point 
over the $j$ line, inside the desymmetrized billiard.

Finally, we can see that $\phi_{j}(x)+\alpha_{j}(x)$ defines the angle 
swept by $Q_{j}(x^{(j)})$ in a continuous way. Hence, 
$\tilde \varphi_{j}= \tilde 
\varphi_{j-1}+\phi_{j-1}(x_{j})+\alpha_{j-1}(x_{j})$
for $j \! \geq \!2$ prevents the phase from jumping when 
changing from one line to the next one.

\section{Resonances on the billiard boundary}

\label{resboundary}

In Section \ref{domain} we described the construction of resonances on 
the billiard domain. This is only suitable for calculations carried out in 
the low energy region, where these wave functions are simple. If we 
include longer orbits as well as higher energy families of the 
shortest ones we need to consider a reduction to a surface of 
section. Also, for calculations of matrix elements and for obtaining explicit 
expressions in terms of classical quantities when $ \hbar \rightarrow 0$ 
is essential to perform this reduction.
As a surface of section we can take any differentiable curve $\xi$ with a 
coordinate  $q$ along it and another $\eta$ orthogonal to $\xi$ at the 
point $q$ ($\eta = 0$ over the curve). Consider an orbit $\gamma$, its 
$l$ crossings with this curve (taken to be at $q_j$ ($j= 1, \ldots , l)$, 
with angles $\theta_j$ on $\xi$ and at $x_j$ on $\gamma$) define a number 
$l$ of $m$th excited wave packets 
as the function $\psi_\gamma^{(m)}$ is considered only over $\xi$. 
This is in fact the representation of $\psi_\gamma^{(m)}$ on the section. 
In the case of billiards, systems that are bounded by rigid walls, we can 
take this boundary as the surface of section, equipped with Birkhoff 
cordinates, i.e., the boundary arclength $q$ and the 
tangential momentum $p$ at the bounce. The origin of $q$ is at the point 
$(X,Y)=(0,R)$ and it grows in the clockwise sense. But for Dirichlet 
boundary conditions $\psi_\gamma$ is null to order $\sqrt{\hbar}$ on this 
surface. Then, we can take 
\begin{equation}
\varphi_{\gamma}^{(m)}(q) \equiv \frac{\partial 
\psi_\gamma^{(m)}}{\partial \eta} (x,y)
\label{normalderivative1}
\end{equation}
as the representation of $\psi_\gamma^{(m)}$ on the section. 
We are going to obtain the general expression for this function. 
In the neighbourhood of a bounce point, $\varphi_\gamma^{(m)}$ is given 
by the combination of 2 terms, one corresponding to the incoming path 
and the other to the outgoing one (see Fig. \ref{bounce1}). 
\begin{figure}[h]
\psyoffset=-14cm
\psxoffset=7cm
\psboxto(7cm;7cm){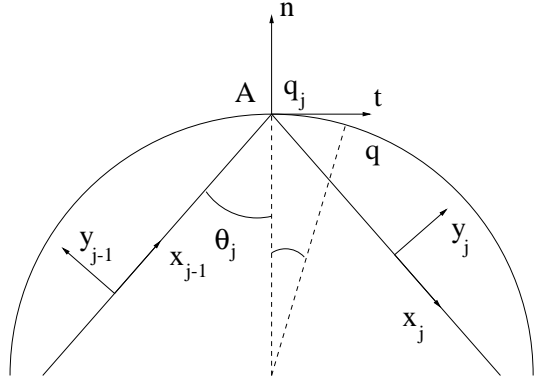}
\caption[]{\label{bounce1} Incoming and outgoing paths of a given 
trajectory at a bounce point on the billiard boundary. Coordinates 
($x_{j-1}$,$y_{j-1}$) and ($x_j$,$y_j$) on the trajectory correspond 
to the incoming and 
outgoing paths; $t$,$n$ are the tangential and normal coordinates to 
the boundary at the bounce point $A$ ($A$ is the origin of $x_j$ and 
$n$). Finally $q$ is the arclength coordinate, with value $q=q_j$ at the 
$j$th bounce.}
\end{figure}

From this Figure we see that the normal $n$ (the former general $\eta$ 
coordinate specialized for this case) and tangential $t$ 
coordinates can be related to the trajectory coordinates by means of 
simple rotations (being $x_j= \sin{(\theta_j)}t - \cos{(\theta_j)}n$, 
$y_j= \cos{(\theta_j)}t + \sin{(\theta_j)}n$ and $x_{j-1}= \sin{(\theta_j)}t+ 
\cos{(\theta_j)}n$, $y_{j-1}= - \cos{(\theta_j)}t + \sin{(\theta_j)}n$). 
We can consider $t \simeq R((q-q_j)/R - (q-q_j)^3/(6R^3))$ for the 
tangential coordinate, where $R$ is the curvature at the bounce. 
Also, for the normal and tangential coordinates  to the boundary at 
the bounce point, the relation $n \simeq -t^2/(2R)$ is 
valid (over the boundary). For this reason we are going to take 
$y \simeq \cos (\theta) t \simeq \cos (\theta) (q-q_j)$ and $x \simeq 
\sin (\theta) (q-q_j)$ in expressions on the boundary at the 
lowest order in $\hbar$. As stated above, we can write the general 
expression for resonances near the $j$th bounce as a sum of 
two contributions of the type described by Eq (\ref{excita}), 
arising from the incoming and outgoing paths, respectively. 
We can write a real expression for the normal 
derivative to the lowest order in $\hbar$ 
(we remember here Eq (\ref{realres}) 
and that the sin function vanishes to order $\sqrt{\hbar}$ 
on the boundary): 
\begin{eqnarray}
\varphi_j^{(m)}(q) & = & - k \; \cos{\theta_j} \; 
\left( \frac{kR}{|Q(x_j)|^2} 
\right)^{1/4} \; f^{(m)}(\xi) \;  \nonumber \\ 
 &   & \frac{2}{\sqrt{L}} \; \cos{[k \cos^2{\theta_j} 
(q-q_j)^2 g_j(x_j) +  k \sin \theta_j} (q-q_j) + \Delta]
\label{normalderivative2}
\end{eqnarray}
where $f^{(m)}(\xi)$ is the same defined after Eq (\ref{realres}) 
in Section 2, but now taking $\xi=\sqrt{kR} (q-q_j) 
\cos \theta_j / |Q(x_j)|$ and $\Delta=k x_j - N_b(x_j^{+}) \pi/2  
- (m + 1/2) \Phi_j(x_j) - F_0 - m G_0$.

As previously mentioned we have evaluated the expression on the 
boundary by means of the $\sqrt{\hbar}$ order approximation in the 
$x$ and $y$ variables, this being a linear approximation in the 
boundary variable $q$. Note that the ordering of bounces and lines 
is as follows: $q_1$ be the first bounce on the boundary of the 
stadium, then $j \; = \; 1$ is the outgoing line from it; this 
process extends up to $L/2$. It is important to mention that if two 
lines reach the bounce point, then an additional factor of 2 is 
needed in order to account for the different weight that these 
points have in ``self-retracting orbits (librations)'' with respect 
to turning points; in the case of rotations the overall factor 
can be considered part of the normalization condition. 

We want to obtain expressions for resonances on the boundary of the 
first quarter of the billiard. Symmetry properties must be 
taken into account and one way to do so is by considering 
``contributions'' coming from bounces that lie outside 
the first quarter of the boundary (see Fig. \ref{desym}). 
The domain counterpart of 
these bounces are taken into account by Eq (\ref{ec2}) of Section 2. 
In fact lines are continued outside the desymmetrised domain. 
But regarding boundary expressions we must explicitely introduce 
them. It is clear that in the semiclassical limit these 
contributions will tend to zero, but for finite $\hbar$ we 
must add to the previous expressions the same ones but 
evaluated at $q_j' = -q_j$; $\theta_j' = -\theta_j$ when 
$q_j \neq 0$, and $q_j' = L/2 - q_j$; $\theta_j'=-\theta_j$ 
when $q_j \neq L/4$ (see Fig. \ref{desym}). The case ``c'' 
of Fig. \ref{desym} can be disregarded.

\begin{figure}[h]
\psyoffset=-14cm
\psxoffset=3cm
\psboxto(9cm;9cm){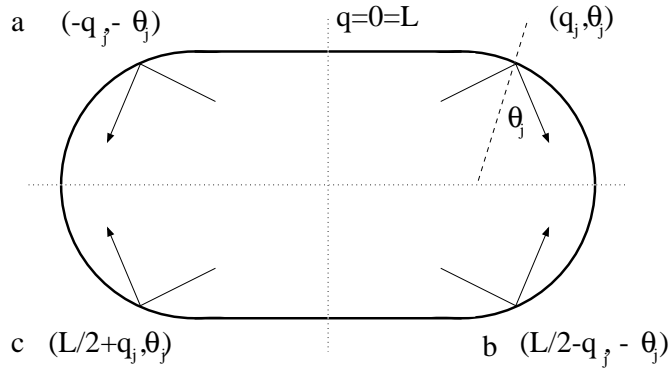}
\caption[]{\label{desym} Horizontal and vertical symmetry related 
bounces and corresponding values of their Birkhoff coordinates 
on the stadium boundary.}
\end{figure}

Then, by adding up all the contributions at each bounce along 
the trajectory we arrive at the complete expression for the 
resonance on the boundary. This procedure is analogous to the 
one devised to construct the resonance on the domain by adding 
all contributions coming from different paths. For completeness 
we are going to define the hyperbolic part of 
the Hamiltonian acting over a resonance on the boundary as: 
\begin{eqnarray}
  &  & \frac{\partial(\hat H - E)/(\hbar^2/2 {\rm m})}{\partial n} \; 
\psi_j^{(m)} |_{\xi} \;  \equiv  \hat H_{eff} \varphi_j^{(m)}(q) = 
\nonumber \\
 &  & - s \; k \; f'(x) \; \lambda \; [ \sqrt{(m+1)(m+2)} \;
\varphi_j^{(m+2)} \; + \; \sqrt{(m-1) m} \; \varphi_j^{(m-2)}],
\label{Hbound}
\end{eqnarray}

In the remaining part of this section we introduce corrections to the 
general expression for the resonance over the boundary at each bounce. 
They take into account the curvature of the manifolds that is 
not included in the previous linear approximation. Furthermore, they 
apply in a direct way to the already defined expressions for the 
resonances and the hyperbolic Hamiltonian. 

Eq (\ref{normalderivative2}) describes the normal derivative of 
the resonance on a tangential section to the boundary at a given 
bounce point. But this is not enough if we want to make calculations 
with these expressions. We therefore need to include the way manifolds 
depart from the unstable and stable directions given by the 
vectors $\xi_u$ and $\xi_s$ respectively. 
We will denote the points on the unstable (or the stable) 
manifold corresponding to the fixed point $(q_0,p_0)$ by 
$(\tilde q, \tilde p(\tilde q)) \; = \; (q-q_0, p-p_0)$. A description 
of this situation can bee seen in Fig. \ref{manifold}. In this Figure 
$\tilde p (\tilde q)$ is the function that defines the unstable 
manifold (this procedure applies equally well to the stable manifold). 
In the following we drop the subindex labelling the $j$th bounce 
so that making the notation easier. 
In order to proceed we need to change coordinates from the Birkhoff 
set to the ones defined by the unstable and stable directions. This 
task is done by the matrix ${\cal B}^{-1}$: 
\[
\left( \begin{array}{c}
           u(\tilde q)     \\
           s(\tilde q)   
                \end{array}  \right) = {\cal B}^{-1} 
\left( \begin{array}{c} 
	   \tilde q         \\
	   \tilde p (\tilde q) 
                \end{array}  \right),	   
\]
where ${\cal B}$ is the matrix with columns given by the unstable and 
stable vectors. They are evaluated at ``half a bounce'' and 
projected on the Poincar\'e section. This can be obtained by means of: 
\[
\left( \begin{array}{cc}
           1/\cos \theta & 0     \\
           0 & \cos \theta   
                \end{array}  \right) \;  
\left( \begin{array}{cc} 
	   1 & 0         \\
	   R/\cos \theta & 1 
                \end{array}  \right) \;
 \left( \begin{array}{cc} 
	   a & b         \\
	   c & d 
                \end{array}  \right),	   
\]
where the first matrix projects the vectors on the section and the 
second matrix stands for subtracting half a bounce to the third
one on the right. In this last case, columns are the unstable and 
stable vectors evaluated after the considered bounce 
(see previous Section for details).

\begin{figure}[h]
\psyoffset=-18cm
\psxoffset=8cm
\psboxto(6cm;6cm){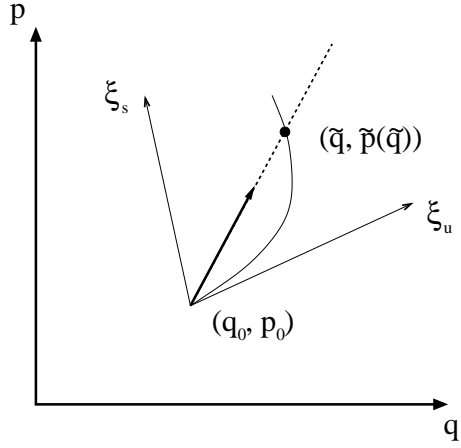}
\caption[]{\label{manifold} Description of point $(\tilde q, 
\tilde p(\tilde q)) \; = \; (q-q_0, p-p_0)$ on the unstable manifold 
corresponding to the fixed point at $(q_0,p_0)$ on the Poincar\'e 
section $(q,p)$.}
\end{figure}

With these functions at hand it is easy to see that the direction 
of the chord that goes through the points $(q_0,p_0)$ and 
$(\tilde q, \tilde p(\tilde q))$ is given by the vector $\xi_u + 
f_u(\tilde q) \xi_s$ with $f_u(\tilde q) \equiv s(\tilde q)/u(\tilde q)= 
\alpha_u \tilde q + \beta_u \tilde q^2 + \ldots$ 
In order to evaluate the effect on the wavefunctions this description 
of the manifolds should be averaged, and we propose a heuristic 
method to do it. A numerically verified 
convenient choice turns out to be the following weighted average:
\begin{equation}
\bar f_u(\tilde q) \equiv \int_0^{\tilde q} f_u(q') 
\frac{\tilde q' dq'}{\tilde q^2 /2} = \frac{2}{3} 
\alpha_u \tilde q + \frac{1}{2} \beta_u \tilde q^2 + \ldots
\label{smooth1}
\end{equation}
Of course, this can also be done for the function $\bar f_s (\tilde q)$ 
which can be expanded in terms of $\tilde q$ as 
\begin{equation}
\bar f_s(\tilde q) = \frac{2}{3} 
\alpha_s \tilde q + \frac{1}{2} \beta_s \tilde q^2 + \ldots
\label{smooth2}
\end{equation}
which gives a complete description of these second order effects. 
Then, the correction to the vectors $\xi_u(\tilde q)$ and 
$\xi_s(\tilde q)$ is the most direct method to include 
them in the resonances. The new vectors now depend on the 
boundary variable $\tilde q$ and must satisfy also the 
normalization condition $\xi_u(\tilde q) \wedge \xi_s(\tilde q) = J$. 
In fact, the normalization factor does not play any role in the  
wave function since, for instance, it cancels out in $\Gamma(\tilde q)$. 
Therefore, we do not show it in the following formulae for the 
$\tilde q$ dependent unstable and stable vectors: 
\begin{equation}
\xi_u(\tilde q) = (1 + \eta \bar f_{u})
[\xi_u + \bar f_u(\tilde q)  \xi_s]
, \; {\rm and} \;  
\xi_s(\tilde q) = (1 + \eta \bar f_{s}) 
[\xi_s + \bar f_s(\tilde q) \xi_u],
\end{equation}
where $\eta=-sgn (q_u q_s) 
\sin^2 \theta$. The factors $(1 + \eta \bar f_{u})$ and 
$(1 + \eta \bar f_{s})$ are due to the fact that, for high values of 
$p$, projecting the unstable and stable vectors on the Poincar\'e 
section results into a stretching 
of them in $q$ direction and a contraction in $p$. 
Then, when modified to consider nonlinear 
effects by means of $\bar f_u (\tilde q)$ and $\bar f_s (\tilde q)$, 
a similar definition for both vectors is guaranteed by this factor.
Finally, the way in which all this corrections enter the previous 
expressions is by new versions of $Re[P(\tilde q)/Q(\tilde q)]$ and 
$\phi(\tilde q)$ expanded in powers of $\tilde q$. 
It is worth to mention that these $Q$ and $P$ are different from those 
obtained in the previous section, the new ones are on the surface of section. 
Then, the final expressions can be obtained by taking into account that now, 
for instance, $Q=(q_u+ \bar f_u(\tilde q) \; q_s) \; (1 + \eta \; 
\bar f_u(\tilde q)) + \; i \; (q_s + \bar f_s(\tilde q) \; q_u ) \; 
(1 + \eta \; \bar f_s(\tilde q))$ and equivalently for $P$. 
Then, it is easy to see that 
\begin{equation}
\tilde g (\tilde q) = Re \left(\frac{P}{2Q} (\tilde q) \right) = 
\frac{p_u q_u + p_s q_s}{2(q_u^2 + q_s^2)} 
+ \frac{g_q \; R \; \tilde q}{3(q_u^2+q_s^2)^2}+ {\cal O}({\tilde q}^2), 
\label{expansion1}
\end{equation}
with $g_q=(\alpha_u + \alpha_s) (q_u^2-q_s^2) + 2 
(\alpha_u - \alpha_s) |q_u q_s| \sin^2 \theta$, 
and where the $\tilde q$ independent part can be related to the expression 
for $g(x)$ given previously (see Section 2). In that case expressions 
were valid in the domain but here they are given on the boundary (notice 
the different coordinate dependence).
Finally, 
\begin{equation}
\tilde \phi=\phi + \frac{2}{3} [(\alpha_s q_u^2 - \alpha_u q_s^2) + 
(\alpha_u - \alpha_s) |q_u q_s| \sin^2 \theta]
 \frac{\tilde q}{q_u^2+q_s^2}+ {\cal O}({\tilde q}^2).
\label{expansion2}
\end{equation} 
Then, the expression for the resonance on the boundary (at each bounce) 
becomes: 
\begin{eqnarray}
\varphi^{(m)}(\tilde q) & = & - k \; \cos{\theta} \; 
\left( \frac{kR}{|Q_0|^2} 
\right)^{1/4} \; f^{(m)}(\xi) \;  \nonumber \\ 
 &   & \frac{2}{\sqrt{L}} \; \cos [k  
 \tilde g( \tilde q) \tilde q^2 +  k \sin (\theta) \tilde q + 
 \tilde \Delta]
\label{normalderivative3}
\end{eqnarray}
with $\tilde \Delta$ the same as in Eq (\ref{normalderivative2}) but 
taking $\tilde \phi$ in place of $\phi$ in the expression for 
$\Phi$. As can be seen we have corrected the phase only since the 
amplitude corrections are not relevant. 
To summarise, we have modified the vectors $\xi_u$  and $\xi_s$ 
in order to take into account the effect of the curvature of 
the manifolds in the resonances on the boundary. Our approach 
allows us to keep the expressions in their compact fashion.

\subsection{Scar function: from the domain to the boundary}

\label{scarboundary}

Now that we have obtained a basis of resonances expressed on 
the billiard boundary for a given 
trajectory $\gamma$ and for 
a quantized energy $E_{\gamma}$ (Eq (\ref{normalderivative3})), 
we can translate this to the scar functions. They are linear 
combinations of resonances having the form \cite{ver1}: 
\begin{equation}
\phi_\gamma = \sum_{j=0}^{N} c_j \;\psi_\gamma^{(4j)} /
\sqrt{{\sum }_{j=0}^{N} \;c_{j}^2}, 
\label{scardom}
\end{equation}
with minimum dispersion $\sigma$, where
\begin{equation}
\sigma^2\equiv<\phi_{\gamma}|(\hat{H}-E_{\gamma})^2 | \phi_{\gamma}>
=<\phi_{\gamma}| \hat{H}_{h}^2 | \phi_{\gamma}>.
\label{disper}
\end{equation}
The boundary representation only amounts to take the normal derivative 
in each term in the previous sum. Also, the second order corrections enter 
immediately through each resonance. We illustrate this by means of 
Figs. \ref{orbs} and \ref{funcs}, 
where several examples of short periodic orbits and linear density 
plots of the Husimi distributions of the corresponding scar functions 
are displayed. In this latter case the solid lines that pass through 
each fixed point on the Poincar\'e section represent the unstable 
and stable manifolds. Finally, the relevance of the new formulation 
can be appreciated with the aid of Fig. \ref{linear_ex}, where 
the scar function in the linear approximation can be observed, 
in this case the one corresponding to 
orbit 2 (see Fig. \ref{orbs} for reference). 

\begin{figure}[h]
\psyoffset=-17cm
\psxoffset=5cm
\psboxto(8cm;8cm){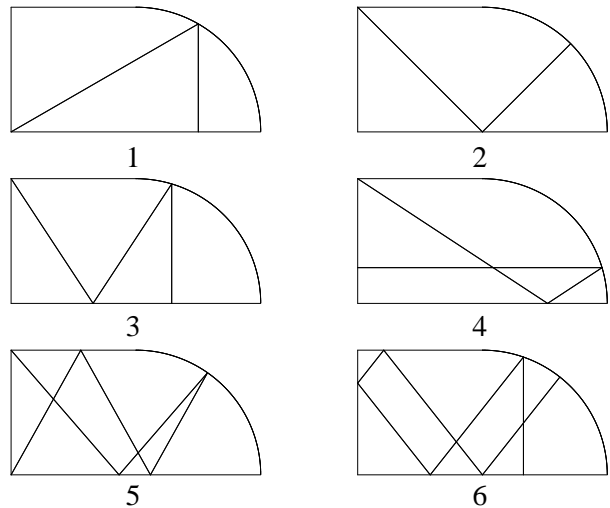}
\caption[]{\label{orbs} Some of the shortest periodic orbits of 
the (desymmetrised) Bunimovich stadium billiard.}
\end{figure}

\begin{figure}[h]
\psyoffset=-27cm
\psxoffset=-3cm
\psboxto(15cm;15cm){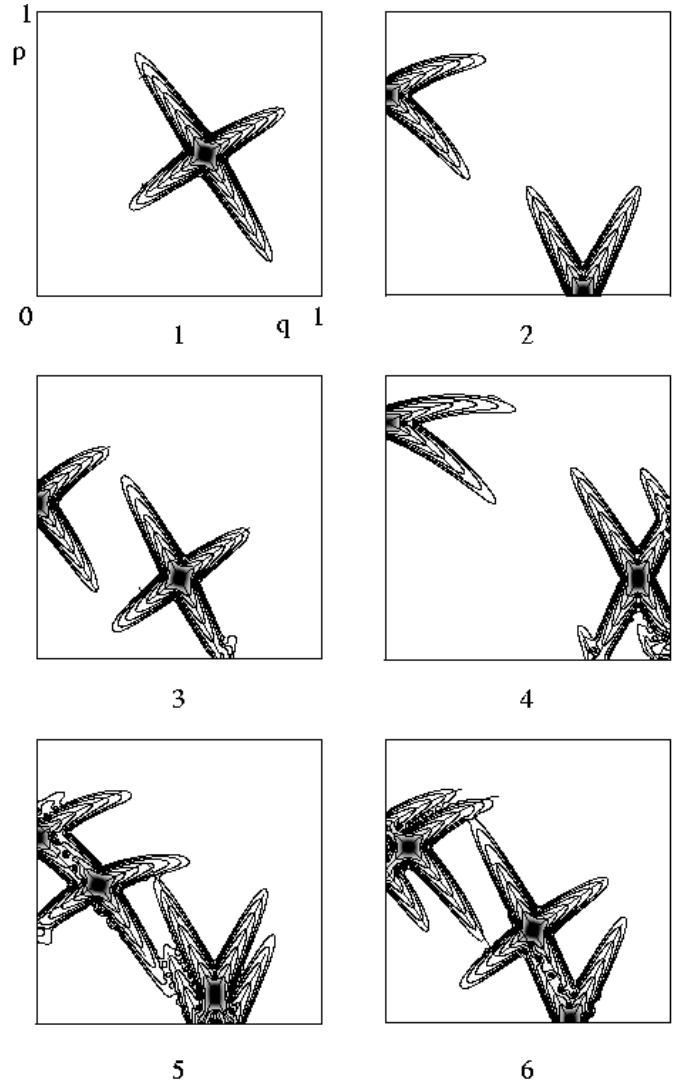}
\caption[]{\label{funcs} Linear density and logarithmic contour plots 
of Husimi distributions 
for the even scar functions corresponding to the orbits displayed in Figure 
\ref{orbs} (numbers below them are the same as for the orbits). 
A logarithmic scale of uniform level ratio $1/e$ from the maximum downwards 
was used in the contour plots. Different levels of gray, uniformly 
distributed, complete this picture. Solid lines passing through fixed 
points represent the unstable and stable manifolds.
The wavenumbers are the nearest to 1000 allowed by the quantisation 
conditions.}
\end{figure}

\begin{figure}[h]
\psyoffset=-19cm
\psxoffset=14cm
\psboxto(5cm;5cm){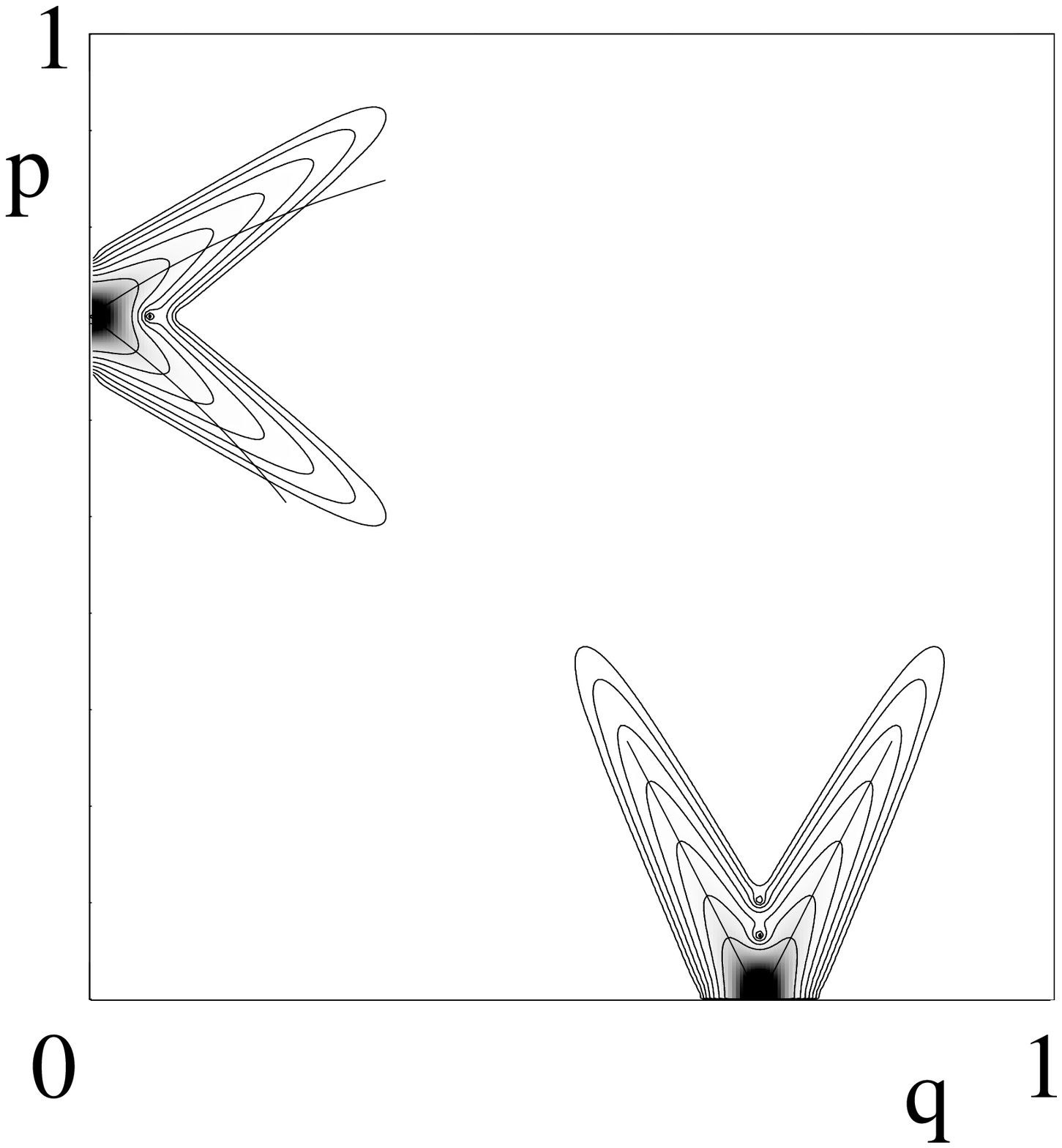}
\caption[]{\label{linear_ex} Linear density and logarithmic contour plot 
of the Husimi distribution 
for the even scar function corresponding to orbit 2 displayed in Figure 
\ref{orbs}. The same scales and details of Fig. \ref{funcs} have been 
considered here.}
\end{figure}

For comparison purposes, we would like to show 
the scar functions on the domain also. The same set of orbits is used for 
constructing them. They can be seen in  
Figure \ref{domainf}.

\begin{figure}[h]
\psyoffset=-15cm
\psxoffset=-2cm
\psboxto(15cm;15cm){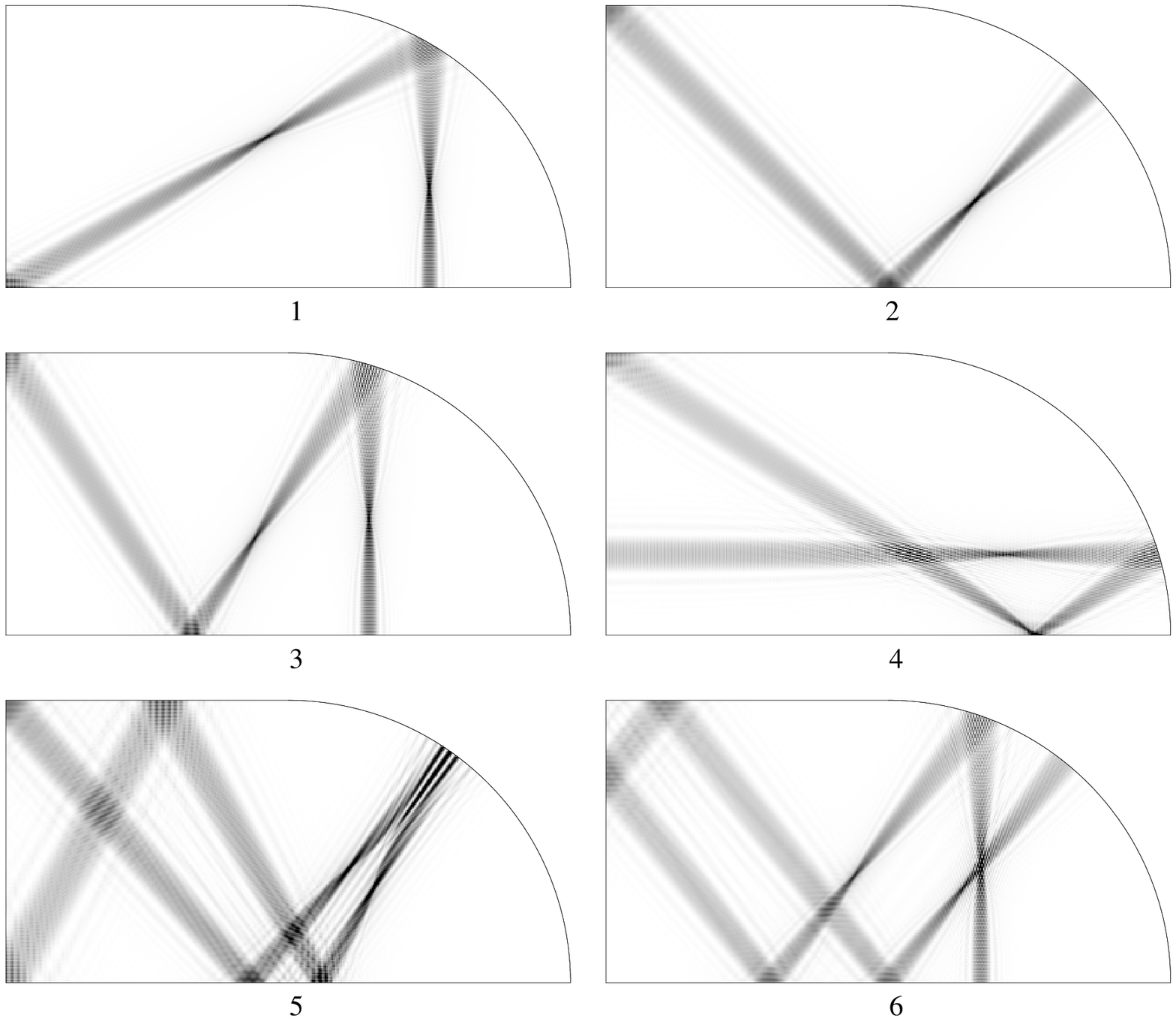}
\caption[]{\label{domainf} Linear density plots of the scar functions of 
Figure \ref{funcs} on the domain of the desymmetrised stadium billiard.}
\end{figure}

\section{Conclusions}

\label{Conclusions}

Resonances on periodic orbits form a convenient basis for the 
investigation of chaotic eigenfunctions.
In the present paper we have extended the construction of resonances 
without excitations in the Bunimovich stadium billiard domain to 
resonances with transversal excitations and also to scar functions. 
A detailed explanation is given, providing an explicit local 
expression for these wave functions.
We have also extended the construction of resonances and scar functions 
from the domain to the boundary of this billiard. 
This procedure enables different sorts of applications, the most 
remarkable one being the possibility to extend calculations of 
eigenstates (in the context of the semiclassical theory of short 
periodic orbits \cite{ver2,ver3}) well above the first low-lying ones. 
In principle this task could have been carried out on the domain also, 
but it would have been very demanding in terms of numerical 
effort and practically turns out to be impossible. The construction 
of resonances and scar functions on the boundary is one of the main 
results of this paper.

We have also accounted for the departure of the unstable and stable 
manifolds from the linear regime, which is the other important main 
result of the 
present work. The departure is reflected in modified expressions 
for the resonances and scar functions. 
The developed method is of general scope and not only can be applied to 
general systems but also to any kind of non linear behaviour of 
the unstable and stable manifolds. This is important to underline, since 
the previous construction of resonances was related only to the linear 
motion around the trajectories, and its area of approximate validity 
shrinks as the energy grows. With the improved but still compact 
expressions for the resonances and scar functions we are able to work 
with a constant effective area.   

Moreover, the direct geometrical approach involved in accounting for 
second order effects due to the projection on the boundary Poincar\'e 
section makes the expressions extremely suitable for calculations. 
A deep exploration of a great number of highly excited 
eigenfunctions arises as a realistic possibility due to the improved 
speed inherent to one dimensional calculations. This will allow us to 
study large sets of eigenfunctions and evaluate statistical measures. 

Finally, with these expressions several integrals made on the domain 
will turn out to be one dimensional Gaussian integrals now, simplifying 
the theoretical analyses.

\section*{Acknowledgments} 

This work was partially supported by SECYT-ECOS. We are grateful to Henning 
Schomerus for his helpful suggestions on the original manuscript.


\end{document}